\documentclass[twoside]{dis07}
\usepackage[latin1]{inputenc}
\usepackage[dvips]{graphicx,epsfig,color}
\usepackage{wrapfig,rotating}
\usepackage{amssymb,amsmath,array}

\newcommand{\lsim}{\raisebox{-4pt}{$\,\stackrel{\textstyle
                                                         <}{\sim}\,$}}

\begin{document}
\title{Electroproduction of longitudinally polarized vector mesons}

\author{Peter Kroll
%
\thanks{The author thanks Markus Diehl for presenting this talk at the  
DIS07 conference.}
%
\vspace{.3cm}\\
%
Universit\"at Wuppertal,  Fachbereich Physik\\
D-42097 Wuppertal, Germany
%
}

\maketitle

\begin{abstract}
It is reported on an analysis of electroproduction of light vector
mesons at small Bjorken-$x$ ($x_{\rm Bj}$) within the handbag
approach. The  partonic subprocesses, meson electroproduction off 
quarks or gluons, are calculated within the modified perturbative 
approach (m.p.a.) in which quark transverse momenta are retained. The soft 
hadronic matrix elements, generalized parton distributions (GPDs), 
are constructed from the CTEQ6 parton distribution functions (PDFs) by means 
of double distributions. The cross sections for longitudinal polarized
virtual photons evaluated from this approach, are 
in very good agreement with experiment in a wide range of kinematics. 
\end{abstract}

It has been shown~\cite{rad96} that, at large photon  virtuality
$Q^2$, meson electroproduction factorizes in partonic subprocesses, 
electroproduction off gluons or quarks, $\gamma^* g(q)\to M g(q)$, and
GPDs, representing soft proton matrix elements. It has also been shown
that the dominant amplitude is that for transitions from longitudinally 
polarized virtual photons ($\gamma^*_L$) to like-wise polarized vector 
mesons ($V_L$). Other transitions are suppressed by inverse 
powers of $Q^2$. In the following it is reported on an 
analysis~\cite{second} of the process $\gamma^*_L p\to V_L p$,
within this handbag factorization scheme carried through in the 
kinematical regime of low $x_{\rm Bj}$ ($\lsim 0.2$).  

The mentioned two partonic subprocesses lead to the following
contributions to the amplitude ${\cal M}_V$ for the process 
$\gamma^*_L p \to V_Lp$: ($x_g=0, x_q=-1$)
\begin{equation} 
{\cal M}^{g(q)}_V \,=\, e \sum_a e_a {\cal C}^a_V \int^1_{x_{g(q)}}
   dx\, {\cal H}_V^{g(q)}(x,\xi,Q^2,t=0) \,H^{g(a)}(x,\xi,t)\,,
\label{ampli}
\end{equation} 
which are to be summed coherently. The sum in (\ref{ampli}) runs over
all quark flavors while $e_a$ denotes the corresponding quark charges
in units of the positron charge $e$. For $\rho$ and $\phi$ production
the non-zero flavor weight factors ${\cal C}^a_V$ read
${\cal C}^u_\rho\,=\,-{\cal C}^d_\rho\,=\,1/\sqrt{2}$ and 
${\cal C}^s_\phi\,=\,1$, respectively. The amplitude (\ref{ampli}) 
refers to proton helicity non-flip, the flip amplitude is neglected 
because our interest lies in small $-t$. The functions $H^{g(a)}$ 
represent GPDs for gluons and quarks. They are functions of three 
variables - $t$, a momentum fraction $x$ and skewness $\xi$. The
latter one is kinematically fixed by 
$\xi\simeq x_{\rm Bj}/(2-x_{\rm Bj})[1+m_V^2/Q^2]$ in a small 
$x_{\rm Bj}$ approximation ($m_V$ being the mass of the vector meson).
To vector meson electroproduction the GPD $\widetilde{H}$ and 
$\widetilde{E}$ do not contribute while $E$ can be ignored in the
region of small $x_{\rm Bj}$ since it contributes $\propto \xi^2$ to  
the proton helicity non-flip amplitude. Thus, only the GPD $H$ is  
required. In (\ref{ampli}) only its $t$-dependence  
is considered. That of the subprocess amplitude ${\cal H}_V$ provides 
power corrections of order $t/Q^2$ which are neglected. In the GPDs 
$t$ is scaled by a soft parameter, actually the slope of the
diffraction peak.

The GPDs are constructed with the help of double distributions 
\cite{muller}. The chief advantage of this construction is the 
guaranteed polynomiality of the GPDs and the correct forward limit 
$\xi, t\to 0$. As is well-known at low $x$ the PDFs behave as powers 
$\delta_i$ of $x$. These powers are determined by the intercepts of
appropriate Regge trajectories. The Regge behaviour of the PDFs is
transferred to the GPDs at $t=0$ by means of the double distribution 
construction. It seems plausible to generate the $t$-dependence of the
GPDs by Regge ideas, too. Assuming linear Regge trajectories 
$\alpha_i(t)=\alpha_i(0) +\alpha_i't$ ($i=$ g, sea, valence) and 
exponential $t$-dependencies of the Regge residues, one may employ the
following ansatz for the double distribution ($n_g=n_{\rm sea}=2$,
$n_{\rm val}=1$)
\begin{equation} 
f_i(\beta,\eta,t)\,=\,{\rm e}^{b_it}\,|\beta|^{-\alpha_i't}\,h_i(\beta)
\,\frac{\Gamma(2n_i+2)}{2^{2n_i+1}\Gamma^2(n_i+1)}\,
\frac{[(1-|\beta|)^2-\eta^2]^{n_i}}{(1-|\beta|)^{2n_i+1}}\,, \nonumber
\end{equation}
where the function $h_i$ represents a PDF suitably continued to negative 
values of $x$. The GPDs are then obtained by an integral over $f_i$
\begin{equation}
H^{i}(x,\xi,t)\, = \, \int_{-1}^1 d\beta\,\int_{-1+|\beta|}^{1-|\beta|}
              d\eta\, \delta(\beta+\xi\eta-x)\,f_i(\beta,\eta,t)\,. \nonumber
\end{equation}

In Ref.\ \cite{second} the Regge parameters are fixed in the following
way: The integrated cross section $\sigma_L \sim \int dt |{\cal M}_V|^2$
behaves $\propto W^{\delta_g(Q^2)}$ at fixed $Q^2$ and small $x_{\rm Bj}$.
Thus, $\delta_g$ can be fixed from the HERA data
\cite{h1-99,zeus-98,zeus-05}. A fit provides 
$\delta_g=0.10+0.06 \ln{(Q^2/4\,{\rm GeV^2})}$. For the slope of the
gluon trajectory the value $\alpha_g'=0.15\,{\rm GeV}^{-2}$ is taken. 
Since the sea quarks mix with the gluons under evolution,
$\alpha_{\rm sea}(t)=\alpha_g(t)$ is assumed. For the valence quarks, 
on the other hand, a standard Regge trajectory is taken - 
$\alpha_{\rm val} = 0.48+0.90\,{\rm GeV}^{-2}t$. The slope parameter of 
the gluon and sea quark Regge residue is obtained from a fit to the HERA data 
on the differential cross section \cite{h1-99,zeus-05}: 
$b_g\,=\,b_{\rm sea}\,=\, 2.58\,{\rm GeV}^{-2} +0.25\,{\rm GeV}^{-2}\,
\ln{[m^2/(Q^2+m^2)]}$ 
($m$ being the proton's mass). In the zero skewness limit the valence quark 
GPDs read  
\begin{equation}
H^q_{\rm val}(x,\xi=0,t=0)\,=\, {\rm e}^{b_{\rm val}t}\,q_{\rm val}(x)\,.\nonumber
\end{equation}
This is very close to the ansatz advocated for in Ref.\  \cite{DFJK4}   
in order to extract the zero-skewness GPDs from the nucleon form factor data.
The comparison with that analysis reveals that one may choose $b_{\rm val}=0$.  

\begin{wrapfigure}{r}{0.5\columnwidth}
\centerline{\includegraphics[width=0.45\columnwidth, bb=34 312 535 700]
{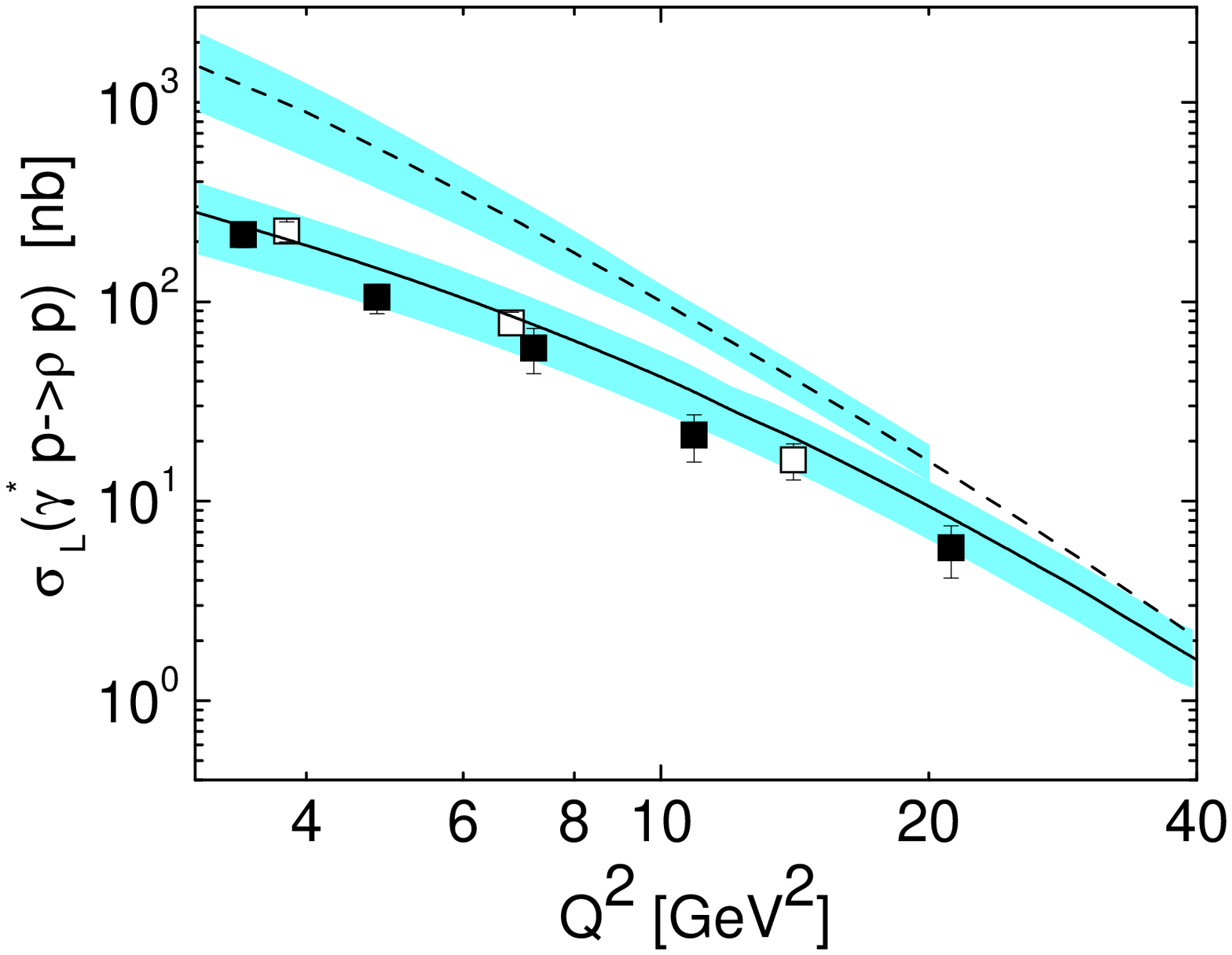}}
\caption{$\sigma_L$ for $\rho$ production at $W=75\,{\rm GeV}$. Data taken 
from H1 \cite{h1-99} (solid squares) and ZEUS \cite{zeus-98} (open squares). 
The solid (dashed) line represents the handbag result using the m.p.a.\ 
(collinear appr.). The error bands are due to 
the uncertainties of the PDFs.}
\label{Fig:1} \vspace*{0.03\textwidth}
\centerline{\includegraphics[width=0.45\columnwidth, bb=40 325 530 700]
{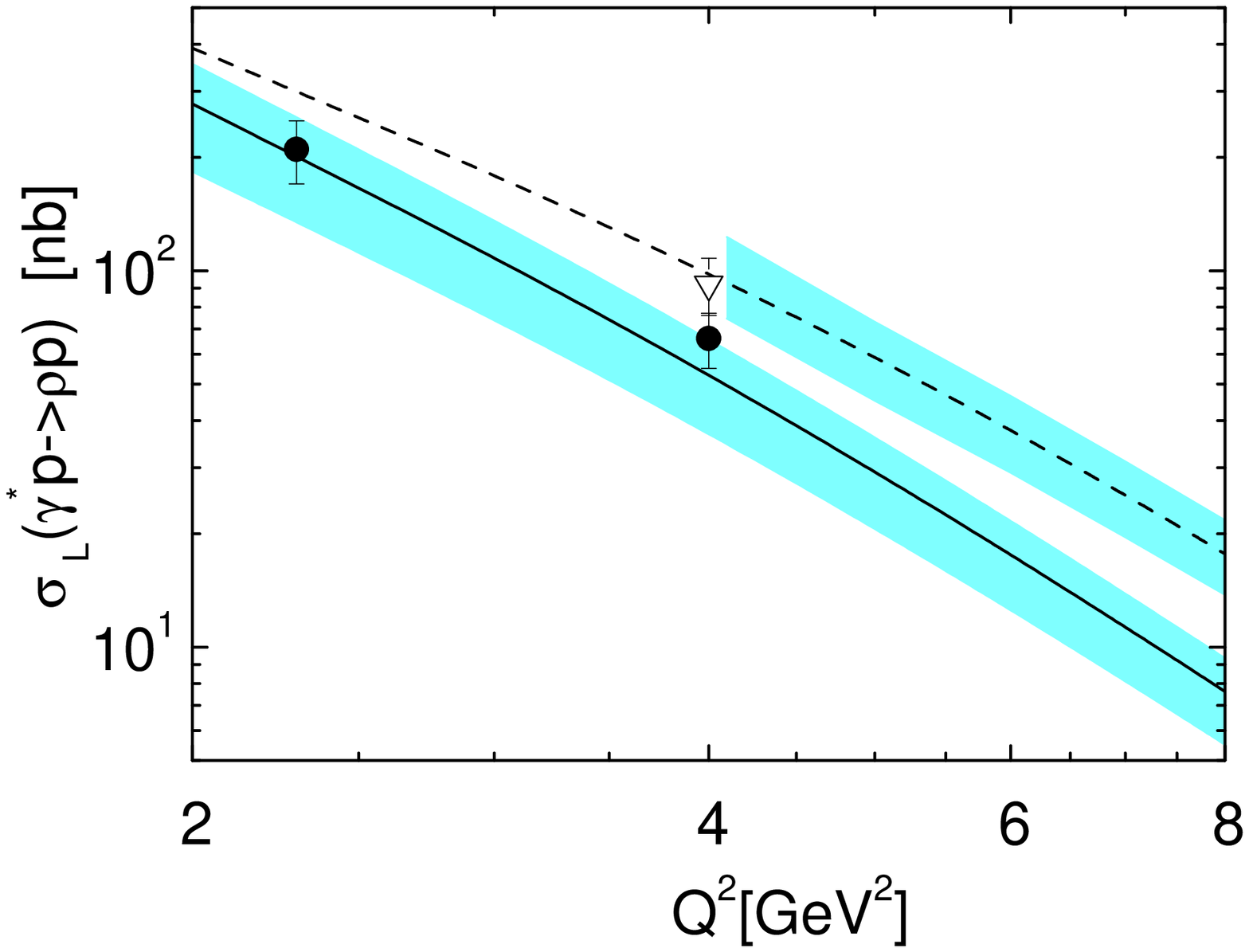}}
\caption{$\sigma_L(\rho)$ versus $Q^2$ at $W=5\,{\rm GeV}$ (solid line) and 
$10\,{\rm GeV}$ (dashed line). Data taken from HERMES \cite{hermes} and
E665 \cite{e665}.}  
\label{Fig:2}
\end{wrapfigure}

Working out the subprocess amplitudes from the relevant Feynman graphs 
in collinear approximation and to LO, one arrives at the following amplitude for 
$\rho$ production
\begin{equation}
{\cal M}_\rho \,=\, e \frac{8\pi\alpha_s}{N_cQ}f_\rho\,
 \langle 1/\tau \rangle_\rho\, \frac1{\sqrt{2}}\,
\Big\{\frac1{2\xi}I_g + \kappa_s C_F I_{\rm sea} + \frac13C_FI^u_{\rm val}
+\frac16C_F I^d_{\rm val}\Big\}\,.
\label{structure}
\end{equation}
The integral $I_g$ reads
\begin{equation}
I_g=2\int_0^1 dx \frac{\xi H^g(x,\xi,t)}{(x+\xi)(x-\xi+i\epsilon)}\,. \nonumber
\end{equation}
Analogous expressions hold for the other two integrals. For $\phi$
production the decay constant $f_\rho$ and the $1/\tau$ moment of 
the $\rho$ distribution amplitude are to be replaced by the
corresponding quantities for the $\phi$ meson. The charge factor
$1/\sqrt{2}$ is to be replaced by $-1/3$ and there is no valence 
quark contribution. For simplicity it is assumed that the $u$ and 
$d$ sea quark GPDs are proportional to that of the strange quark. 
The factor of proportionality, $\kappa_s$, is obtained from the CTEQ6 
PDFs. It is about 2 at $Q^2=4\,{\rm GeV}^2$ and tends towards 1 for 
increasing $Q^2$. Evaluating the GPDs from the CTEQ6M PDFs \cite{cteq} 
and adopting the asymptotic $\rho$ meson distribution amplitude,
leading to $\langle 1/\tau\rangle=3$, one can work out the cross 
section $\sigma_L$ for $\rho$ production. The result, shown in Fig.\ 
\ref{Fig:1}, is evidently too large by order of magnitude at low
$Q^2$. The deviations diminish with increasing $Q^2$. Note that there
are large NLO corrections \cite{ivanov} which cancel the LO term to a 
large extent. Wether the inclusion of higher orders lead to agreement 
with experiment is unknown as yet.    
 
As is well-known from studies of the large momentum transfer behaviour
of electromagnetic form factors, the collinear approximation becomes 
inconsistent in the end-point regions, since the contributions from large 
transverse separations, ${\bf b}$, of the quark and antiquark forming 
the meson are not sufficiently suppressed. In order to eliminate that 
defect the so-called m.p.a.\ has been invented \cite{sterman} in which 
quark transverse degrees of freedom are retained and the accompanying 
gluon radiation ( a Sudakov factor) is taken into account. This
m.p.a.\ is employed in the calculation of the partonic subprocesses. 
Instead of distribution amplitudes rather meson wave functions have to 
be used in this approach. Actually, a Gaussian one 
$\sim {\rm exp}[-a_V^2 k^2_\perp/(\tau(1-\tau))]$ is utilized in 
\cite{second}. The transverse size parameters are considered as free 
parameters to be adjusted to the experiment ($a_\rho=0.75\,{\rm GeV}^{-1}$, 
$a_\phi=0.70\,{\rm GeV}^{-1}$). The structure of the amplitude as given
in Eq.\ (\ref{structure}) still holds if the m.p.a.\ is used, even the 
relative strength of the various contributions remain to be the same. 

\begin{wrapfigure}{r}{0.5\columnwidth}
\centerline{\includegraphics[width=0.45\columnwidth, bb=32 330 501 743]
{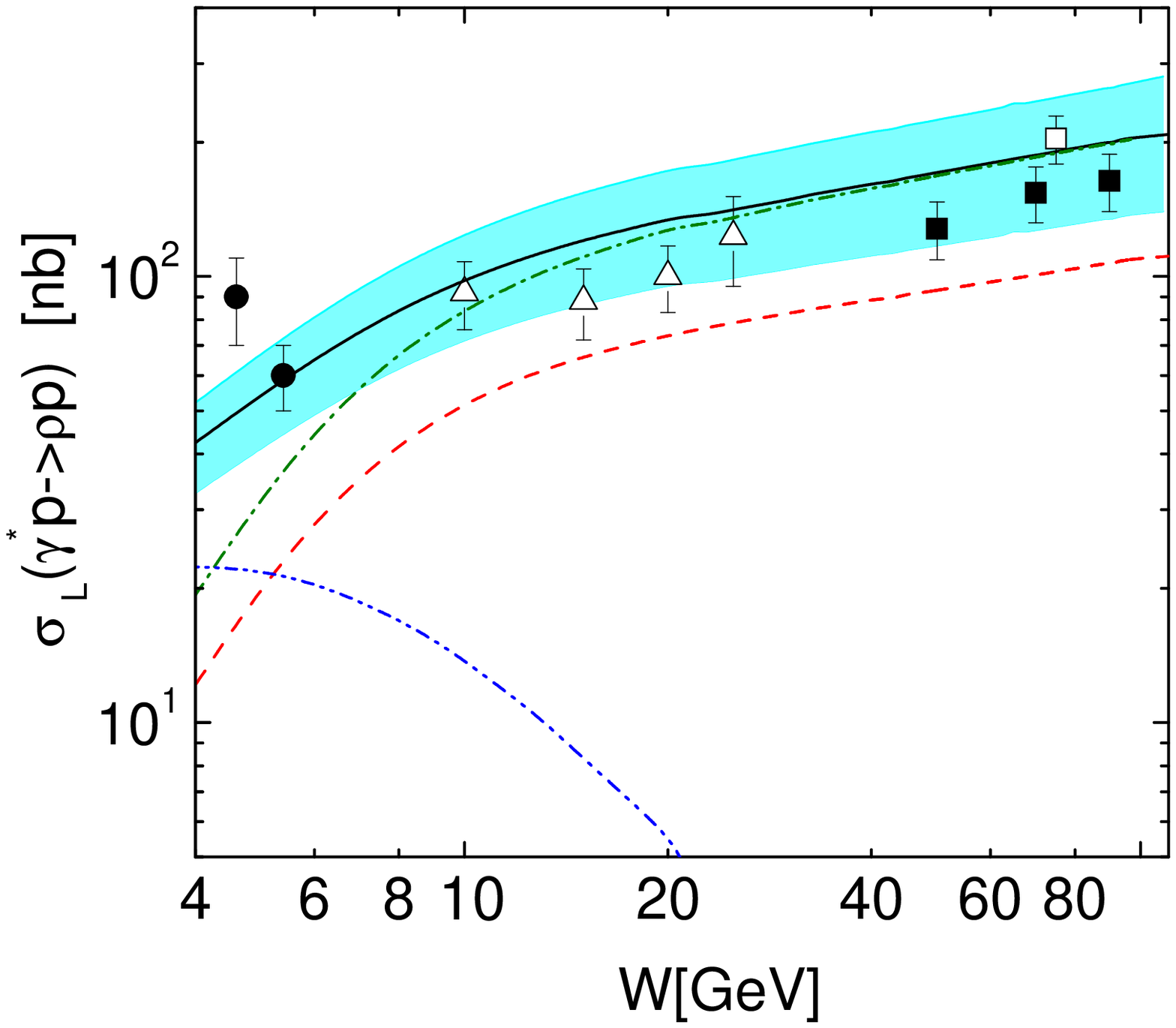}} 
\caption{$\sigma_L(\rho)$ at $Q^2=4\,{\rm GeV}^2$. Data taken from
H1 \cite{h1-99} (solid square), ZEUS \cite{zeus-98} (open square), E665 \cite{e665}
(open triangle) and HERMES \cite{hermes} (solid circle). The dashed (dash-dotted, 
dash-dot-dotted) line represents the gluon (gluon+sea, (gluon+sea)-valence 
interference plus valence quark) contribution. For further notation, cf. Fig.\ 
\ref{Fig:1}.} 
\label{Fig:3} \vspace*{0.09\textwidth}
\centerline{\includegraphics[width=0.45\columnwidth, bb=10 292 580 701]
{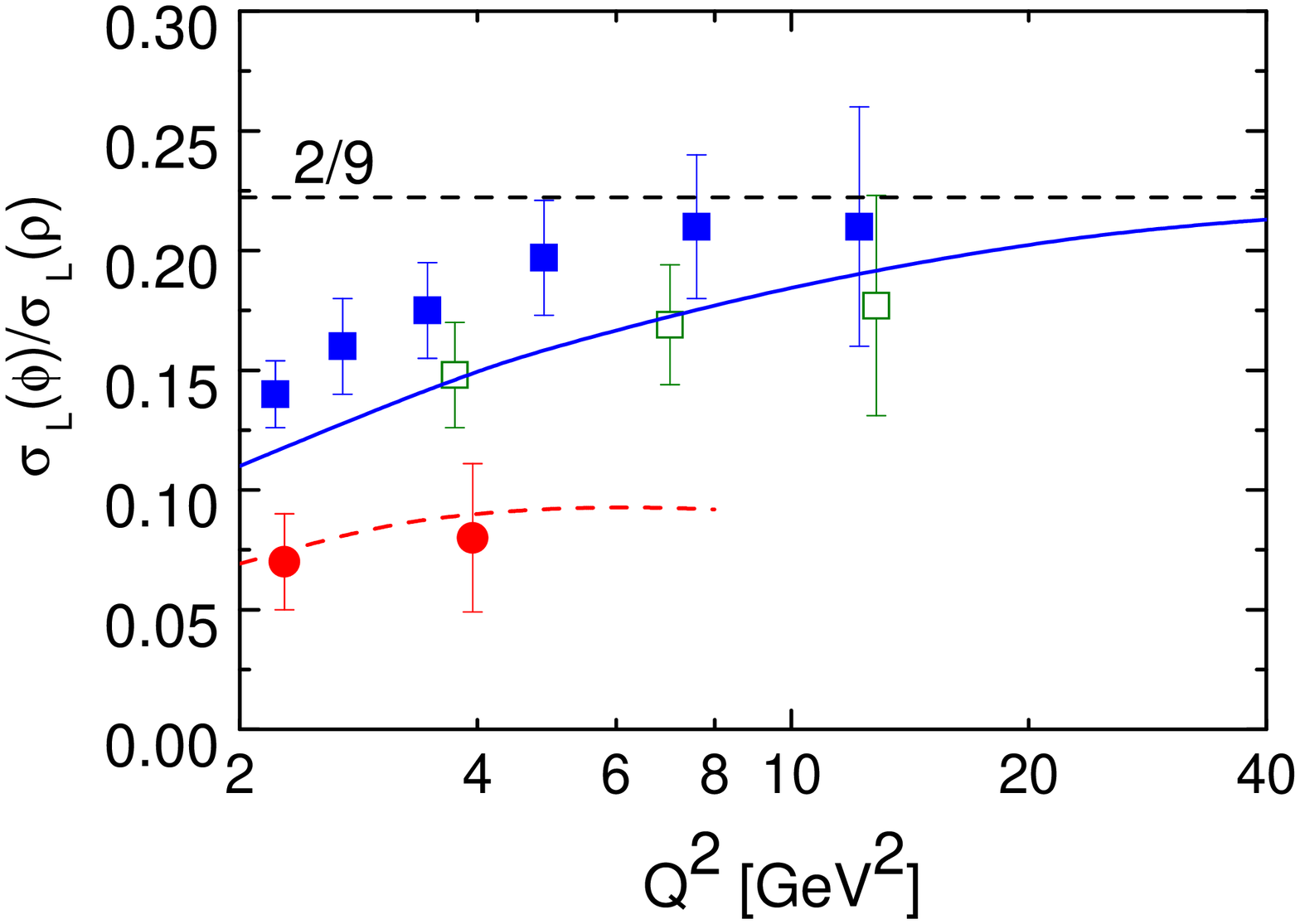}}
\caption{The ratio $\sigma_L(\phi)/\sigma_L(\rho)$. The solid (dashed) line
represents the handbag result at $W=75\, (5)\, {\rm GeV}$. Data are taken 
from H1, ZEUS and HERMES.}
\label{Fig:4}
\end{wrapfigure}

Detailed comparison with experiment is made in Ref.\ \cite{second}. 
Here, only a few results are presented. As one can see from  Fig.\ 
\ref{Fig:1} if the m.p.a.\ is used, the handbag result for 
$\sigma_L(\rho)$ is in fair agreement with the HERA data 
\cite{h1-99,zeus-98} in particular if one considers the uncertainties 
in the theoretical results due to the errors of the PDFs. Results for 
$\sigma_L$ at lower values of $W$ are shown in Fig.\ \ref{Fig:2}. 
Next, in Fig.\ \ref{Fig:3} the energy dependence of $\sigma_L(\rho)$ 
is displayed. The figure also reveals the prominent role of the gluonic 
contribution. The valence quark contribution is only significant below 
$10\,{\rm GeV}$. Results of similar quality  are obtained for $\phi$ 
production. Here, only the ratio of the cross section for $\phi$ and 
$\rho$ production is shown in Fig.\ \ref{Fig:4}. For $Q^2\to \infty$ 
the handbag approach predicts 
$\sigma_L(\phi)/\sigma_L(\rho) \longrightarrow 2/9 (f_\phi/f_\rho)^2\,=\,0.248$.
The deviations from this limit seen in Fig. \ref{Fig:4} at finite
$Q^2$, are generated by the breaking of flavor symmetry in the sea and, 
although to a lesser extent, by the meson wave function. The low value
of the ratio at $W=5\,{\rm GeV}$ is due to the additional valence quark contribution
to the $\rho$ cross section.

In summary - the handbag factorization scheme with the partonic subprocesses 
calculated within the m.p.a.\ and GPDs constructed from the double distributions 
provides reasonable results for the 
longitudinal cross section of $\rho$ and $\phi$ electroproduction in a large
range of $Q^2$ 

\noindent
and $W$. As shown in Ref.\ \cite{first} this approach can also 
be applied to the amplitude for transversally polarized photons. The infrared 
singularities occuring for that amplitude in collinear approximations are 
regularized in the m.p.a.\ by the quark transverse momenta.
In Ref.\ \cite{first} this amplitude has been calculated  for HERA kinematics
assuming that only the gluonic subprocess contributes. The inclusion of quarks
for this amplitude is in progress. This analysis will provide results on 
$\sigma_T$ and on various spin density matrix elements for the same range of $Q^2$ 
and $W$ as for $\sigma_L$.


\begin{footnotesize}


\end{footnotesize}



\begin{thebibliography}{99}
\bibitem{rad96} A.V.\ Radyushkin, 
Phys.\ Lett. {\bf B385}, 333 (1996);
J.C.\ Collins {\it et al.}, 
Phys.\ Rev.\ {D56}, 2982 (1997).

\bibitem{second} S.~V.~Goloskokov and P.~Kroll,
  Eur.\ Phys.\ J.\  C {\bf 50}, 829 (2007).


\bibitem{muller} D.\ M\"uller {\it et al},
Fortschr. Phys. {\bf 42}, 101 (1994);
A.V.\ Radyushkin, 
Phys.\ Lett. {\bf B449}, 81 (1999). 


\bibitem{h1-99} C.\ Adloff {\it et al}., [H1 collaboration],
Eur.\ Phys.\ J.\ {\bf C13}, 371 (2000). 

\bibitem{zeus-98} J.\ Breitweg {\it et al}., [ZEUS collaboration],
Eur.\ Phys.\ J.\ {\bf C6}, 603 (1999).

\bibitem{zeus-05} S.\ Chekanov {\it et al}., [ZEUS collaboration],
 Nucl.\ Phys.\ {\bf 718}, 3 (2005). 

\bibitem{DFJK4} M.~Diehl, T.~Feldmann, R.~Jakob and P.~Kroll,
 Eur. Phys. J. {\bf C39}, 1 (2005).

\bibitem{cteq} J. Pumplin {\it et al}, JHEP {\bf 0207}, 012 (2002).


\bibitem{ivanov} D.~Y.~Ivanov {\it et al.}
  JETP Lett.\  {\bf 80}, 226 (2004)
  [Pisma Zh.\ Eksp.\ Teor.\ Fiz.\  {\bf 80}, 255 (2004)], M.\ Diehl
  und W.\ Kugler, these proceedings.
  

\bibitem{sterman} J.\ Botts and G.\ Sterman, 
Nucl.\ Phys.\ {\bf B325}, 62 (1989).

\bibitem{hermes} A.\ Airapetain {\it et al.}, [HERMES collaboration],
Eur.\ Phys.\ J.\ {\bf C17}, 389 (2000).

\bibitem{e665} M.R.\ Adams {\it et al.}, [E665 collaboration],
Z.\ Phys.\ {\bf C74}, 237 (1997).

\bibitem{first} S.V.\ Goloskokov and P. Kroll, 
Eur. Phys. J.\\ {\bf C42} 281 (2005).
\end{thebibliography}
\end{document}